# A Convolutional Neural Network Based Approach to Recognize Bangla Spoken Digits from Speech Signal

Ovishake Sen
*Computer Science and Engineering*
*Khulna University of Engineering*
*& Technology, Khulna, Bangladesh*
*sen1607066@stud.kuet.ac.bd*

Al-Mahmud
*Computer Science and Engineering*
*Khulna University of Engineering*
*& Technology, Khulna, Bangladesh*
*mahmud@cse.kuet.ac.bd*

Pias Roy
*Computer Science and Engineering*
*Khulna University of Engineering*
*& Technology, Khulna, Bangladesh*
*roy1607071@stud.kuet.ac.bd*

*Abstract*—Speech recognition is a technique that converts human speech signals into text or words or in any form that can be easily understood by computers or other machines. There have been a few studies on Bangla digit recognition systems, the majority of which used small datasets with few variations in genders, ages, dialects, and other variables. Audio recordings of Bangladeshi people of various genders, ages, and dialects were used to create a large speech dataset of spoken '০-৯' Bangla digits in this study. Here, 400 noisy and noise-free samples per digit have been recorded for creating the dataset. Mel Frequency Cepstrum Coefficients (MFCCs) have been utilized for extracting meaningful features from the raw speech data. Then, to detect Bangla numeral digits, Convolutional Neural Networks (CNNs) were utilized. The suggested technique recognizes '০-৯' Bangla spoken digits with 97.1% accuracy throughout the whole dataset. The efficiency of the model was also assessed using 10-fold cross-validation, which yielded a 96.7% accuracy.

*Keywords*—Bangla speech recognition, Bangla spoken '০-৯' digits classification, CNN, MFCCs, Cross-validation

## I. Introduction

Software for speech recognition uses voice data to connect with computers and to understand the data as a user interface. All individuals, from companies to people, use technology broadly. In recent years the technology of speech recognition has grown increasingly common. Software for spoken recognition can create papers during less than half the time for typing. The saving of time owing to greater efficiency, less red tape, numerous activities and visibility of workflow allows better management of priorities, turnarounds, etc. Speech recognition systems are influenced by the speaker's age, gender, quantity, house, fashion, and other characteristics. As a result, there is no guarantee that a strategy that works well for one language would work well for another [1], [2].

## II. Literature Review

In comparison to English or other rich languages, there have been few studies on Bangla isolated and continuous speech recognition. Paul et al. [3] presented a Bangla speech recognition system utilizing pre-emphasis filtering, speech coding, LPC, and ANN. Sultana et al. [4] developed a technique for converting Bangla speech to text using SAPI [5]. Hasnat et al. [6] developed a strategy for constructing an isolated and continuous Bangla voice recognition system applying HMM toolkit. Ahammad et al. [7] suggested a connected digit recognition system employing Backpropagation neural network. They got an average of 89.87% accuracy recognizing 0 to 9 Bangla spoken digits. Using a double-layered LSTM-RNN approach, Nahid et al. [8] developed a technique for constructing a Bangla Speech Recognition system, with 28.7% and 13.2% phon and word detection error rate, respectively. Using HMM CMU Sphinx and Android TTS API, Ahmed et al. [9] presented a voice input speech output calculator and they got 86.7% accuracy for word recognition. To recognize Bangla's short speech instructions, Sumon et al. [10] presented three CNN architectures: MFCC-based CNN model, raw CNN model and pre-trained CNN model utilizing transfer learning where they found 74%, 71% and 73% accuracy,respectively. Islam et al. [11] developed a CNN-based speech recognition system and an RNN-based approach to identify Bengali character level probabilities. Shuvo et al. [2] suggested a Bangla digit recognition system from speech signals using MFCC and CNN, with 93.65% test set recognition accuracy. Sharmina et al. [12] presented an in-depth learning method for identifying Bengali spoken digits using MFCC and CNN, with 98.37% accuracy, 98.37% precision, 98.37% recall, and 98.37% F1-score. Paul et al. [13] suggested a Bangla digit recognition system from speech signals using MFCC and GMM, and their self-built Bangla numeral data set produced 91.7% correct predictions.

## III. Preparatory

### A. Convolutional Neural Network (CNN)

CNN's are a kind of neural network to analyze structured data arrays like pictures. Convolution and subsampling are the two primary operations in the convolutional neural network. The input feature map is convoluted with trainable filters and bias values is subsequently applied during the convolution phase. The convoluted feature

map(CFM) is calculated using Eq(1). During the sub-sampling phase, certain processes like as max and average pooling are conducted in a specified CFM pooling area. Sub-sampled feature map(SFM) is calculated using Eq(2).

$$CFM_{x,y} = f(b + \sum_{i=1}^{k_h}\sum_{j=1}^{k_w} K_{i,j} \circledast I_{x+i,y+j}) \quad (1)$$

Here, f()=activation function, b=Bias, kh,kw=size of kernels and ⊛=two dimensional convolution.

$$SFM_{x,y} = d(\sum_{i=1}^{R}\sum_{j=1}^{C} CFM_{i,j}) \quad (2)$$

Here, d(.)=Sub-sampling operation on a pooling area; R,C= Pooling area size.

### B. Mel Frequency Cepstral Coefficients (MFCCs)

Mel Frequency Cepstral Coefficients (MFCCs) are a type of feature extractor used in voice recognition systems. Using the mel scale, the frequency band is split into sub-bands, and by applying Discrete Cosine Transform(DCT) the cepstral coefficients are extracted. MFCCs may better mimic the human auditory system's reaction. Figure 1 shows general steps to calculate MFCCs. [14]

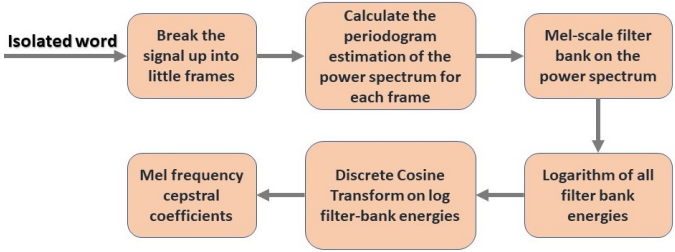

Fig. 1. Processing steps of MFCCs

## IV. Proposed Method

In this paper, a Bangla numeral voice recognition system has been proposed using deep learning models trained on self-developed speech corpus of Bangla digits from '০-৯'. Figure 2 displays the fundamental flow schematic of the proposed CNN-based numerical speech recognition system.

### A. Dataset Construction

As a lack of available public dataset for Bangla spoken digits, a speech dataset of '০-৯' digits of 19 speakers from different locations of Bangladesh has been developed. Gathering voice recordings from people of diverse Bengali dialects, genders, and ages helped to diversify this dataset. Around 400 audio samples were recorded in both noisy and quiet environments for each digit. So, a total of 4,000 audio samples were recorded for Bangla '০-৯' digits. Each audio sample was recorded with a microphone using the 'audacity' recording software, with a sampling rate of 44kHz and 'wav' audio format. Table I illustrates the speakers specification for building the dataset.

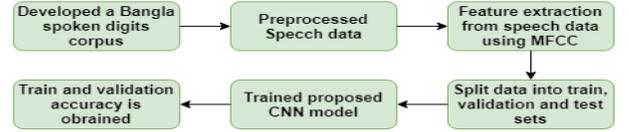
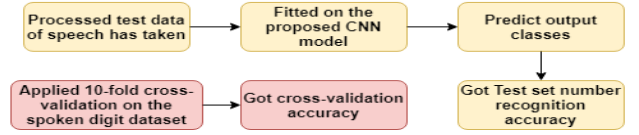

Fig. 2. Fundamental flow schematic of the proposed CNN-based numerical speech recognition system

TABLE I
Speakers specification for building the dataset

| Location of Bangladesh | No. of speakers |
|---|---|
| Bagerhat | 3 |
| Chittagong | 2 |
| Dhaka | 2 |
| Dinajpur | 1 |
| Mymensingh | 1 |
| Narayanganj | 4 |
| Rajshahi | 1 |
| Sylhet | 1 |
| Thakurgaon | 4 |
| Total | 19 |

### B. Preprocessing and Augmentation

Using the 'audacity' recording software, the raw voice recordings were reduced to obtain isolated digit speech. The raw voice data was subjected to audio augmentation techniques such as time shift, speed tuning, background noise mixing, and volume adjustment [15] . The 'noise reduce' and 'pyDub' libraries in Python were used to eliminate noise and silence from the raw voice, respectively.

### C. Feature Extraction and Train Test Split

The basic purpose of feature extraction is to convert the speech waveform into a unique parametric representation from which relevant features can be extracted. In this study, characteristics are extracted using the Mel Frequency Cepstrum Coefficient (MFCC). Using Python's 'Librosa' module, the MFCC features were extracted from the numeral speech data and stored in an array. The voice data was then labeled as '০-৯' depending on the speech data. To recognize numerical Bangla talks, the first 13 MFCC coefficients were employed, together with their first and second delta values, forming a total of 39 mfcc coefficients. The spectrogram of total 39 mfccs coefficients for "০" digit is depicted in figure 3. From the whole dataset, 64%, 16% and 20% of data are used for training purposes, and testing purposes, respectively. So, a total of 2560 audio samples are used for training, a total of 640 audio samples are used for validating, and a total of 800 audio samples are used for testing the proposed model. For this train-test split, the Python's scikit-learn library was employed.

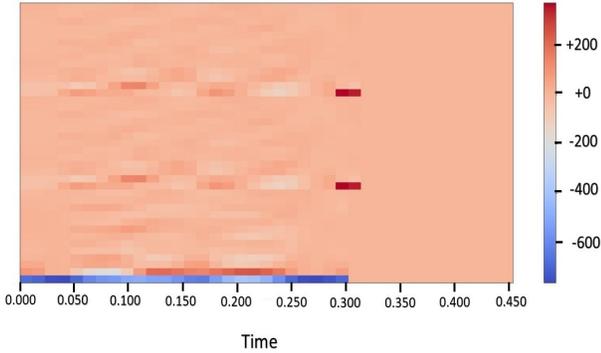

Fig. 3. Spectrogram of 39 mfcc coefficients for "১" digit

### D. Feature Learning and Classification using CNN

CNN architecture was utilized for feature learning and classification purposes. Figure 4 shows the proposed CNN Model Architecture.

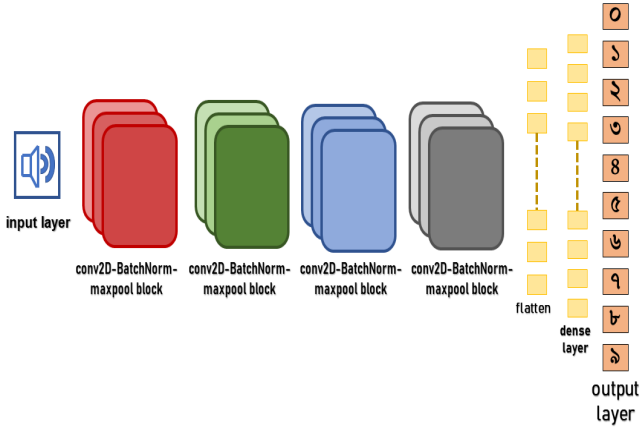

Fig. 4. Proposed CNN Model Architecture

(39*39*1) is the form of the CNN model's input. The preprocessed train data was put into a 2D convolutional layer with a filter size of 24 and a kernel size of (3, 3). This input layer also used the relu activation function and an L2 regularizer with a regularization factor of 0.1. The BatchNormalization layer's output was then transferred to the MaxPooling layer, which had a pool size of (2, 2). The method of combining Conv2D-BatchNormalization-MaxPooling layers was repeated three times, with filter sizes of 32, 64, and 128 in each Conv2D layer.

The output vector is converted to a one-dimensional vector by the flatten layer. The 1D vector is then connected to a 128-unit fully connected layer using a relu activation function. A Dropout layer with a dropout rate of 0.2 was utilized to avoid overfitting. Finally, the softmax layer is employed in the classification process. Categorical cross-entropy and Adam were employed as the loss function and optimizer, respectively. The learning rate is set to 0.0001.

TABLE II
EXPERIMENTS DETAILS AND RESULTS ON PROPOSED CNN MODEL

| Expo no. | Data Size instances | No. of epochs | Batch Size | Learning Rate | Accuracy (%) | Loss | Precision (%) | Recall (%) | F1-Score (%) | Training Time (hour) |
|---|---|---|---|---|---|---|---|---|---|---|
| 1 | 1000 | 1000 | 32 | 0.0001 | 84.80 | 0.96 | 84.80 | 84.80 | 84.80 | 0.20 |
| 2 | 2000 | 1000 | 32 | 0.0001 | 87.25 | 0.81 | 87.25 | 87.25 | 87.25 | 0.75 |
| 3 | 3000 | 1000 | 32 | 0.0001 | 93.33 | 0.35 | 93.33 | 93.33 | 93.33 | 1.75 |
| 4 | 4000 | 1000 | 32 | 0.0001 | 97.1 | 0.52 | 97.1 | 97.1 | 97.1 | 2.5 |

## V. EXPERIMENTAL RESULTS AND ANALYSIS

The proposed approach was evaluated with 4,000 occurrences of the self-created '০-৯' Bangla spoken digits dataset. The model is built with Python, Keras, and Tensorflow. The reserach was carried out using Google Colaboratory, a free cloud-based Jupyter notebook environment. The following formula is used to determine the recognition accuracy of spoken Bangla digits:

$$Accuracy = \frac{Perfectly\ recognized\ digits}{Recognized + Unrecognized\ digits} * 100\% \quad (3)$$

The suggested method was evaluated four times with varied data sizes of 100, 200, 300, and 400 occurrences per digit, yielding accuracy of 84.80%, 87.25%, 93.33%, and 97.1%, respectively. So we can presume that even higher accuracy can be obtained with this model if more data are being fed into this model. The details and outcomes of the four experiments are shown in Table II. Figure 5 shows the improvement of results after providing more data into the proposed CNN model.

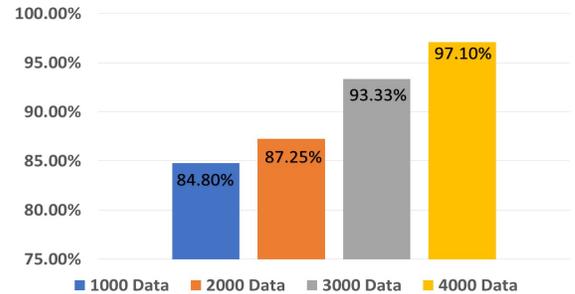

Fig. 5. Improvement of results after increasing the data size

We've noticed individuals from different locations pronounce an arbitrary digit differently. We've also noticed that at the time of pronunciation, some digits are phonetically quite similar to other digits. Such as '৬', '৯' and '৭', '৮' sounds quite similar. Again, '৪' is pronounced differently as 'চারর' and 'চাইরর'. Table III shows the confusion matrix and accuracy, precision, recall, f1-score of each digit. As we can see from the table III, this is why we have found a lesser accuracy in detecting certain digits

TABLE III
Confusion matrix and accuracy, precision, recall, f1-score of each digit

| Digits | ০ | ১ | ২ | ৩ | ৪ | ৫ | ৬ | ৭ | ৮ | ৯ | Accuracy (%) | Precision (%) | Recall (%) | F1-score (%) |
|---|---|---|---|---|---|---|---|---|---|---|---|---|---|---|
| ০ | 78 | 0 | 0 | 0 | 0 | 2 | 0 | 0 | 0 | 0 | 97.50 | 100.00 | 97.50 | 98.73 |
| ১ | 0 | 80 | 0 | 0 | 0 | 0 | 0 | 0 | 0 | 0 | 100.00 | 98.76 | 100.00 | 99.37 |
| ২ | 0 | 0 | 77 | 0 | 0 | 0 | 0 | 0 | 0 | 3 | 96.25 | 96.25 | 96.25 | 96.25 |
| ৩ | 0 | 0 | 1 | 79 | 0 | 0 | 0 | 0 | 0 | 0 | 98.75 | 98.75 | 98.75 | 98.75 |
| ৪ | 0 | 0 | 0 | 1 | 76 | 0 | 1 | 2 | 0 | 0 | 95.00 | 97.43 | 95.00 | 96.20 |
| ৫ | 0 | 0 | 0 | 0 | 0 | 77 | 0 | 0 | 3 | 0 | 96.25 | 97.47 | 96.25 | 96.86 |
| ৬ | 0 | 0 | 0 | 0 | 1 | 0 | 76 | 0 | 1 | 2 | 95.00 | 98.70 | 95.00 | 96.81 |
| ৭ | 0 | 0 | 0 | 0 | 1 | 0 | 0 | 78 | 1 | 0 | 97.50 | 97.50 | 97.50 | 97.50 |
| ৮ | 0 | 1 | 0 | 0 | 0 | 0 | 0 | 2 | 77 | 0 | 96.25 | 92.77 | 96.25 | 94.47 |
| ৯ | 0 | 0 | 0 | 0 | 0 | 1 | 0 | 0 | 0 | 79 | 98.75 | 92.94 | 98.75 | 95.75 |
| Average: | | | | | | | | | | | 97.1 | 97.1 | 97.1 | 97.1 |

TABLE IV
Comparison between proposed and existing approaches

| Article | Dataset summary | Methods | Accuracy on test set |
|---|---|---|---|
| Ahmed et al. [9] | 7 speakers and 2,733 samples | CMU Sphinx and Android TTS API | 86.7% for 0-9 digits |
| Paul et al. [13] | 1,000 samples | MFCC and GMM | 91.7% for 0-9 digits. |
| Ahammad et al. [7] | 30 speakers and Noise-free 260 samples | Segmentation and BPNN | 89.87% for 0-9 digits |
| Nahid et al. [8] | 15 speakers and Noisy 2000 samples | Noise reduction, phoneme mapping and LSTM | For 0-9 digits 28.7% and 13.2% phon and word detection error rate, respectively |
| Shuvo et al. [2] | 120 speakers and Noise free 6,000 samples | CNN | 93.65% for 0-9 digits |
| Sharmina et al. [12] | 5 speakers and 1,230 samples | CNN | 98.37% for 0-9 digits |
| Proposed method | 19 speakers from different gender, age groups and areas of Bangladesh and 4,000 noisy audio samples for '০-৯' Bangla digits | MFCC, CNN and cross-validation | 97.1% for recognizing '০-৯' digits on CNN model and 96.7% for recognizing '০-৯' digits on 10-fold cross-validation |

'৬': 95.00%, '৮': 96.25% and '৪': 95.00% that are spoken ambiguously from different places of Bangladesh. On the contrary, other non-conflicted digits finds quite good accuracy such as, '০': 97.50%, '১': 100.00%, '৩': 98.75% and '৯': 98.75%. The proposed model was also tested using 10-fold cross-validation and got 96.7% accuracy. As we can see the comparison between proposed and existing approaches from table IV that, most existing research activities have been done on either on a smaller dataset or have obtained less accuracy for identifying Bangla spoken '০-৯' digits. In this work, however, we put our suggested model to the test on a large variational dataset of self-created Bangla spoken digits and found that the model performs superbly. We have also performed cross-validation to test the effectiveness and found that it performed admirably.

## VI. CONCLUSIONS

This paper aims to create a convolutional neural network model that can recognize Bangla '০-৯' numerals from the speech input using a self-built Bangla spoken digits dataset. We have created our voice recognition system by collecting Bangla spoken digits, which contain gender, accent, and age criteria. In this dataset, however, we were unable to cover all Bengali dialects. In the future, we want to gather more linguistic data from individuals of all ages and genders in various regions of Bangladesh and will also try to build a better model. This method is also compared to several existing works in the field of classifying '০-৯' digits, indicating its superiority.

## REFERENCES

[1] S. Hossain, M. Rahman, F. Ahmed, and M. Dewan, "Bangla speech synthesis, analysis, and recognition: an overview," *Proc. NCCPB*, 2004.
[2] M. Shuvo, S. A. Shahriyar, and M. Akhand, "Bangla numeral recognition from speech signal using convolutional neural network," in *2019 International Conference on Bangla Speech and Language Processing (ICBSLP)*. IEEE, 2019, pp. 1–4.
[3] A. K. Paul, D. Das, and M. M. Kamal, "Bangla speech recognition system using lpc and ann," in *2009 Seventh International Conference on Advances in pattern recognition*. IEEE, 2009, pp. 171–174.
[4] S. Sultana, M. Akhand, P. K. Das, and M. H. Rahman, "Bangla speech-to-text conversion using sapi," in *2012 International Conference on Computer and Communication Engineering (IC-CCE)*. IEEE, 2012, pp. 385–390.
[5] T. D. Chung, M. Drieberg, M. F. B. Hassan, and A. Khalyasmaa, "End-to-end conversion speed analysis of an fpt. ai-based text-to-speech application," in *2020 IEEE 2nd Global Conference on Life Sciences and Technologies (LifeTech)*. IEEE, 2020, pp. 136–139.
[6] M. Hasnat, J. Mowla, M. Khan *et al.*, "Isolated and continuous bangla speech recognition: implementation, performance and application perspective," 2007.
[7] K. Ahammad and M. M. Rahman, "Connected bangla speech recognition using artificial neural network," *International Journal of Computer Applications*, vol. 149, no. 9, pp. 38–41, 2016.
[8] M. M. H. Nahid, B. Purkaystha, and M. S. Islam, "Bengali speech recognition: A double layered lstm-rnn approach," in *2017 20th International Conference of Computer and Information Technology (ICCIT)*. IEEE, 2017, pp. 1–6.
[9] T. Ahmed, M. F. Wahid, and M. A. Habib, "Implementation of bangla speech recognition in voice input speech output (viso) calculator," in *2018 International Conference on Bangla Speech and Language Processing (ICBSLP)*. IEEE, 2018, pp. 1–5.
[10] S. A. Sumon, J. Chowdhury, S. Debnath, N. Mohammed, and S. Momen, "Bangla short speech commands recognition using convolutional neural networks," in *2018 International Conference on Bangla Speech and Language Processing (ICBSLP)*. IEEE, 2018, pp. 1–6.
[11] J. Islam, M. Mubassira, M. R. Islam, and A. K. Das, "A speech recognition system for bengali language using recurrent neural network," in *2019 IEEE 4th international conference on computer and communication systems (ICCCS)*. IEEE, 2019, pp. 73–76.
[12] R. Sharmin, S. K. Rahut, and M. R. Huq, "Bengali spoken digit classification: A deep learning approach using convolutional neural network," *Procedia Computer Science*, vol. 171, pp. 1381–1388, 2020.
[13] B. Paul, S. Bera, R. Paul, and S. Phadikar, "Bengali spoken numerals recognition by mfcc and gmm technique," in *Advances in Electronics, Communication and Computing*. Springer, 2021, pp. 85–96.
[14] "Mel frequency cepstral coefficient (mfcc) tutorial," accessed: 2021-06-28. [Online]. Available: http://www.practicalcryptography.com/miscellaneous/machine-learning/guide-mel-frequency-cepstral-coefficients-mfccs
[15] D. S. Park, W. Chan, Y. Zhang, C.-C. Chiu, B. Zoph, E. D. Cubuk, and Q. V. Le, "Specaugment: A simple data augmentation method for automatic speech recognition," *arXiv preprint arXiv:1904.08779*, 2019.